\documentclass{sigchi-ext}
\usepackage[T1]{fontenc}
\usepackage{textcomp}
\usepackage[scaled=.92]{helvet} 
\usepackage{float}
\usepackage{graphicx} 
\usepackage{balance}  
\usepackage{booktabs} 
\usepackage{ccicons}  
\usepackage{ragged2e} 

\usepackage{marginnote}
\usepackage[usenames,dvipsnames]{xcolor}
\usepackage{tikz}
\usetikzlibrary{shapes}
\usepackage{amsmath,amssymb}
\usepackage{rotating}

\copyrightinfo{\textbf{AUTHOR BIO:} Megan Strait is a PhD candidate in the Cognitive Science Program at Tufts University.
She received her BA in computer science from Wellesley College in 2010 and her MSc from Tufts University in 2014.
Her graduate research is centered in the areas of human-robot and human-computer interaction.
She is especially interested in the uncanny valley and its implications for the design and use of humanlike robots.
Her dissertation (\emph{Understanding the Uncanny: The Effects of Human Similarity on Aversion towards Humanlike Robots}) advances uncanny valley theory and introduces a novel protocol for evaluating human-robot interactions.}

\title{Conceptualization and Validation of a Novel Protocol for Investigating the Uncanny Valley}

\numberofauthors{1}
\author{%
  \alignauthor{%
    \textbf{Megan K. Strait}\\
    \affaddr{Tufts University} \\
    \affaddr{Medford, MA 02155, USA} \\
    \affaddr{megan.strait@tufts.edu} } }

\def\plaintitle{The Uncanny Valley} \def\plainauthor{First Author, Second Author, Third Author,
  Fourth Author, Fifth Author, Sixth Author}
\def\plainkeywords{Uncanny valley theory, social robotics; human-robot interaction; emotion; emotion regulation}

\hypersetup{%
  pdftitle={\plaintitle}, pdfauthor={\plainauthor},
  pdfkeywords={\plainkeywords}, }

\begin{document}

\maketitle
\RaggedRight{}

\begin{abstract}
Loosely based on principles of similarity-attraction, robots intended for social contexts are being designed with increasing human similarity to facilitate their reception by and communication with human interactants.
However, the observation of an \emph{uncanny valley} -- the phenomenon in which certain humanlike entities provoke dislike instead of liking -- has lead some to caution against this practice.
Substantial evidence supports both of these contrasting perspectives on the design of social technologies.
Yet, owing to both empirical and theoretical inconsistencies, the relationship between anthropomorphic design and people's liking of the technology remains poorly understood.

Here we present three studies which investigate people's explicit ratings of and behavior towards a large sample of real-world robots.
The results show a profound ``valley effect'' on people's \emph{willingness} to interact with humanlike robots, thus highlighting the formidable design challenge the uncanny valley poses for social robotics.
In addition to advancing uncanny valley theory, Studies 2 and 3 contribute and validate a novel laboratory task for objectively measuring people's perceptions of humanlike robots.
\end{abstract}


\section{Introduction}
Loosely based on similarity-attraction theory \cite{EISENBERG1987}, technologies intended for social contexts are being designed with increasing \emph{human similarity} to facilitate their reception by and communication with people.
This anthropomorphization of social technologies represents a particularly powerful mechanism for facilitating human-computer and human-robot interactions.
Specifically, increasing the human similarity of a nonhuman entity can elicit more positive responding from its human interactants, which in turn, leads to positive social outcomes such as improved rapport in application domains such as education, collaboration, and therapy (e.g., \cite{ANDRIST2012}).

\begin{marginfigure}[0pc]
\begin{minipage}{\marginparwidth}
\centering
\includegraphics[width=1\marginparwidth]{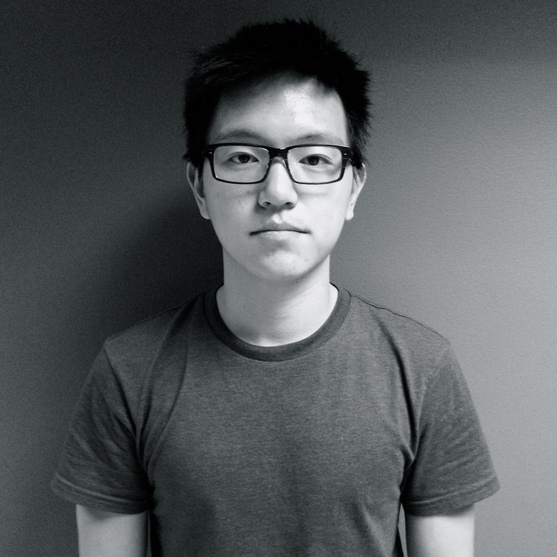}
\includegraphics[width=1\marginparwidth]{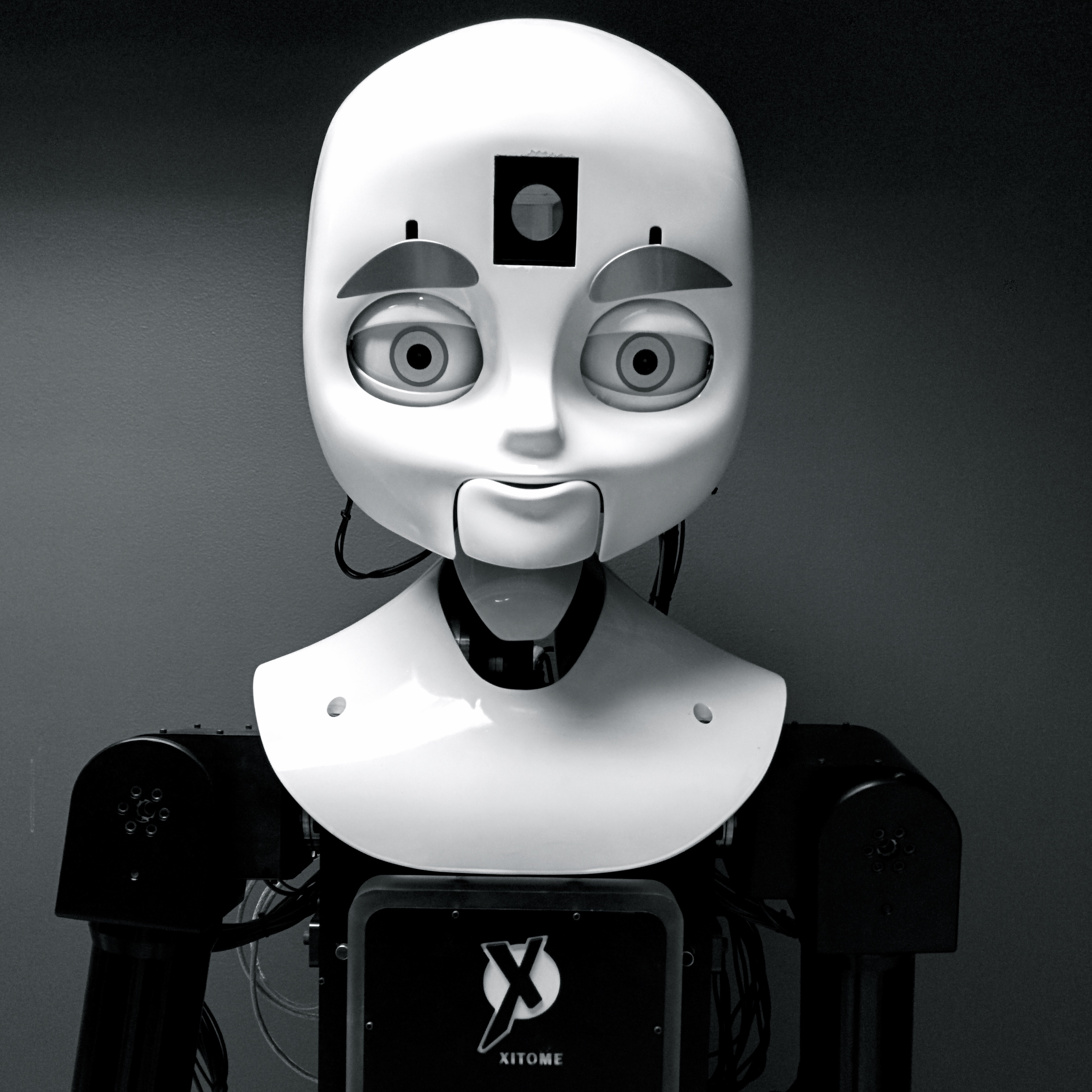}
\includegraphics[width=1\marginparwidth]{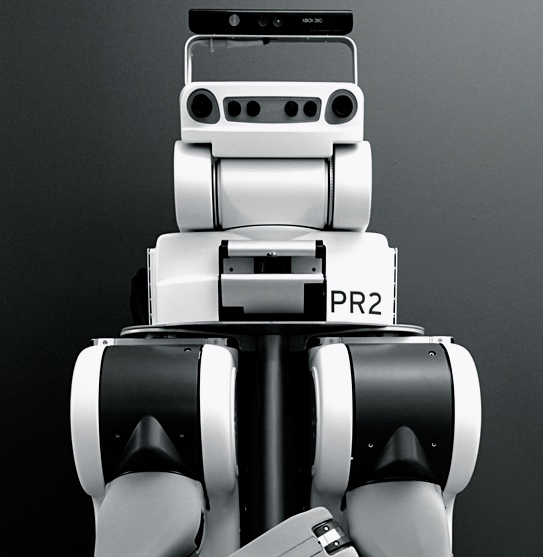}
\caption{Exemplars of a human (top) and two robots of varying human similarity (center, bottom).}
\label{fig:agents}
\end{minipage}
\end{marginfigure}

For example, explicit humanlike cues such as a humanlike face presented on a computer screen (as compared with a text-based computer) lead people to respond more positively to and feel more relaxed with the computer (e.g., \cite{SPROULL1996}), personified interfaces help users engage in tasks (e.g., \cite{KODA1996}), and humanlike web content (socially-rich text and picture elements) increaseses perceptions of usefulness, trust and enjoyment of shopping websites, leading to more favourable consumer attitudes (e.g., \cite{HASSANEIN2007}).
Similarly, equipping a robot with humanlike attributes facilitates the formation of rapport with, empathic responding towards, and positive appraisals of it (e.g., \cite{DUFFY2003, RIEK2009, SAUPPE2015}).

\subsection{The Anthropomorphization of Social Robots}
Within the human-robot interaction community, this has lead to a pervasive assumption that people's liking of social robots is a monotonically-increasing function of human similarity (i.e., greater human similarity is always better).
This assumption is reflected by the sheer number of engineering efforts towards developing humanlike robots (see Table~\ref{tab:humanoids}) to the instantiation of a new field of study (\emph{android science}) devoted to this topic \cite{ISHIGURO2007}.
Consistent with similarity-attraction theory, it is expected that such robots offer more natural and effective interactions by capitalizing on traits which are more familiar and intuitive to people.

This perspective is not unfounded.
In fact, it is rather well-supported by a large empirical base.
Humanlike robots are perceived as more thoughtful (e.g., \cite{ANDRIST2014}, intelligent (e.g., \cite{BROADBENT2013}), and importantly, likeable (e.g., \cite{DUFFY2003, RIEK2009}) relative to their less humanlike counterparts.
People also report greater comfort in their presence (e.g., \cite{SAUPPE2015}) and are more receptive to a robot interlocutor and compliant on collaborative tasks the greater the similarity of the robot (e.g., \cite{ANDRIST2015,PARISE1999}).

\subsection{The Uncanny Valley}
Others have suggested a more nuanced relationship between human similarity and people's liking of anthropomorphic entities.
In particular, the emergence of increasingly humanlike robots and other artificial entities brought to light a competing phenomenon: the \emph{uncanny valley} \cite{MORI1970}.
The uncanny valley refers to the observation of certain entities -- often those with a highly humanlike appearance -- provoking significant discomfort, instead of affinity as would be predicted by similarity-attraction theory.

Over the course of nearly five decades since Masahiro Mori's formal introduction of the uncanny valley theory into scientific discourse, empirical inquiries have compiled substantial evidence of its existence (for a review, see \cite{KATSYRI2015}).
Specifically, people tend to rate highly humanlike entities as eerie and more unnerving than less humanlike instances (e.g., \cite{MACDORMAN2006A}).
Such dislike appears to also manifest in infants \cite{LEWKOWICZ2012,MATSUDA2012} and even other primates \cite{STECKENFINGER2009}, suggesting the general phenomenon is relatively pervasive.

Yet, uncanny valley theory continues to be critically questioned due to various inconsistencies in and shortcomings of empirical probes (e.g., \cite{BARTNECK2009, KATSYRI2015, ZLOTOWSKI2013B}).
For example, people respond negatively towards some but not all instances of highly humanlike agents (e.g., \cite{ROSENTHAL2014A}).
Similarly for certain humanlike attributes, while some find their application provokes dislike (e.g., \cite{PIWEK2014}), others find the exact opposite (e.g., \cite{WANG2006}).
It thus remains difficult to decide which perspective to take in the design of these social technologies.

At the root of the above issues lay several gaps in the literature.
Most notably, the community lacks a consistent methodology for investigating the uncanny valley and its effects.
Moreover, albeit due to practical limitations, the literature largely draws on subjective assessment and on the comparison of very few agents.
Given the high variability of both subjective measures and the appearance of real-world robots, it may serve to explain at least some of the inconsistencies between studies using different robots.

\begin{margintable}[5pt]
\hspace{0mm}
\begin{minipage}{\marginparwidth}
\begin{tabular}{l l}
	\toprule
	Affetto & \cite{ISHIHARA2011}\\
	Aila 	& \cite{LEMBURG2011}\\
	Asimo 	& \cite{SAKAGAMI2002}\\
	BARTHOC & \cite{HACKEL2005}\\
	Baxter 	& \cite{FITZGERALD2013}\\
	CB2 	& \cite{MINATO2007}\\
	Emiew 	& \cite{HOSODA2006}\\
	Geminoid series & \cite{NISHIO2007}\\
	HRP 	& \cite{KANEKO2008}\\
	HRP-4C	& \cite{KANEKO2011}\\
	HUBO series 	& \cite{PARK2007}\\
	iCub 	& \cite{METTA2008}\\
	Kaspar 	& \cite{DAUTENHAHN2009}\\
	Kobian 	& \cite{ZECCA2008}\\
	Kojiro 	& \cite{MIZUUCHI2007}\\
	Enon 	& \cite{KANDA2006}\\
	Flobi 	& \cite{LUTKEBOHLE2010}\\
	MDS 	& \cite{BREAZEAL2008}\\
	Nao 	& \cite{GOUAILLIER2009}\\
	REEM 	& \cite{TELLEZ2008}\\
	Repliee series 	& \cite{MACDORMAN2005}\\
	Robonaut 	& \cite{AMBROSE2000}\\
	Robovie & \cite{ISHIGURO2001}\\
	Saya 	& \cite{HASHIMOTO2006}\\
	Telenoid 	& \cite{OGAWA2011B}\\
	Twendy-One 	& \cite{IWATA2009}\\
	Valkyrie 	& \cite{RADFORD2015}\\
	Wabian 	& \cite{OGURA2006}\\
	Wakamaru 	& \cite{SHIOTANI2006}\\
    \bottomrule
\end{tabular}
\caption{Exemplars of humanlike robots and design series.}
\label{tab:humanoids}
\end{minipage}
\end{margintable}

\subsection{Present Research}
In the following sections, we present three experimental investigations of people's perceptions of a large sample of real-world robots.\footnote{All procedures were approved by the Social, Behavioral, and Educational Research Institutional Review Board at Tufts University. All participants provided written, informed consent prior to participating and received either $\$10/hour$ or course credit as compensation.}
Here we showed participants a series of pictures depicting humans and robots of varying human similarity (low, moderate, and high).
We collected participants' subjective ratings of the agents' appearances on two dimensions: \emph{human similarity} as a manipulation check, and \emph{eeriness} to determine whether the set of humanlike robots refelcts an uncanny valley.

To move towards more objective assessment, Studies 2 and 3 additionally propose and validate a novel protocol for measuring people's behavior towards humanlike robots.
Specifically, Studies 2--3 investigate the link between the uncanny valley and avoidant behavior using the process model of emotion regulation \cite{GROSS1998} as theoretical grounding.
The motivations for doing so follow from the literature on avoidant behavior, defining it as a person's unwillingness to experience negative emotions and their desire to change the form or frequency of situations giving rise to those experiences (e.g., \cite{ELLIOT2001}) -- which is an example of emotion regulation through situation selection.

The implications of avoidant behavior is particularly important to human-computer and human-robot interaction given a primary aim of design is to \emph{facilitate} interaction.
Thus, while increasing a robot's human similarity can effect positive social outcomes, it remains crucial to understand when/why a design effects \emph{negative} responding.
Hence, the purpose of Studies 2--3 was to investigate whether the uncanny valley presents a serious consideration for human-robot interaction via more objective assessment of its impact on people's behavior.
That is, we wanted to determine whether highly humanlike robots can be so emotionally motivating that they evoke avoidant behavior.

In Studies 2--3, we again presented the series of pictures depicting humans and robots of varying human similarity, but with the addition of an option \emph{to press a button to remove the picture} if the participant wished to stop looking at it.
In addition to the subjective ratings of the agents' appearances, we collected the percentage of button presses to measure the frequency of attempts to end encounters with the various agents as an index of avoidant behavior.

\section{Study 1}
Based on Mori's uncanny valley theory, as well as supporting evidence (e.g., \cite{MACDORMAN2005}), we hypothesized that people would rate highly humanlike robots as more eerie than than less humanlike robots and humans (\emph{Hypothesis 1}).
To test this prediction, we conducted a fully within-subjects study in which we manipulated the shown agents' \emph{human similarity}.

\textbf{Materials \& Methods}\\
\emph{Protocol.}
Participants viewed a set of color pictures, each depicting a distinct (robotic or human) agent.
The pictures were obtained from various academic and internet sources, and were categorized into four levels of relative human similarity: robots of \emph{low}, \emph{moderate}\footnote{For brevity, the results and contrast regarding the category of robots with moderate human similarity are excluded in the remainder of the paper.}, or \emph{high} similarity, as well as \emph{human} agents (see Figure~\ref{fig:agents}).
To generalize beyond the appearance of any one agent, we included $15$ instances per category for a total of $60$ pictures.
Each picture was shown for 10-seconds in duration, followed by two prompts for the participant to rate the depicted agent.

\begin{marginfigure}[0pc]
\begin{minipage}{\marginparwidth}
\centering
\includegraphics[width=3.35\marginparwidth, angle=90]{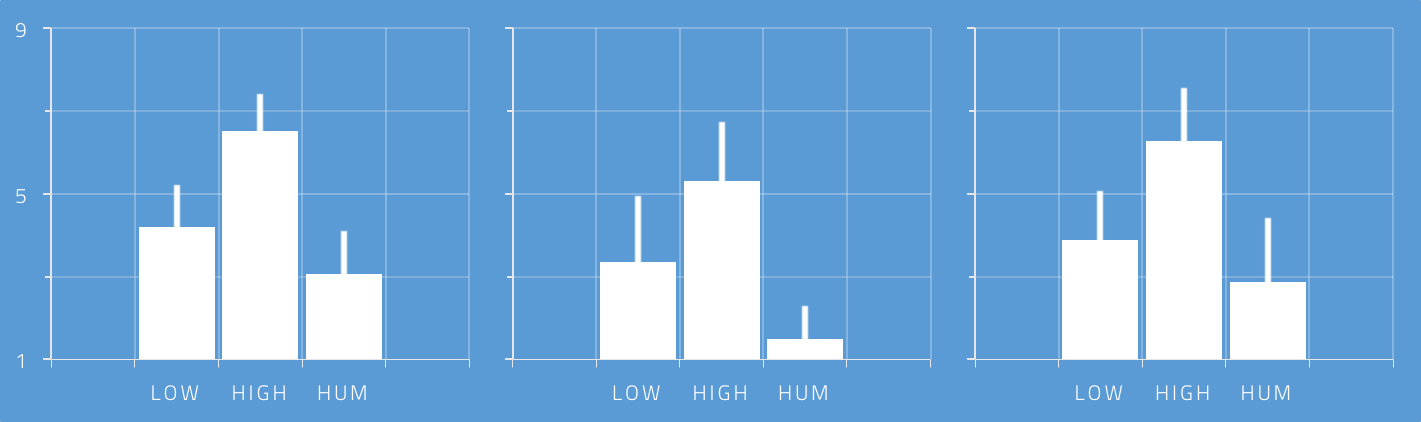}
\caption{Mean eeriness rating by agent category, across Studies 1 (left), 2 (middle), and 3 (right).}
\label{fig:eeriness}
\end{minipage}
\end{marginfigure}

\emph{Measures.}
We collected participants' subjective ratings of the depicted agents' appearances on two dimensions: \emph{human similarity} and \emph{eeriness}.
As a fully within-subjects design was used, the ratings were averaged (by participant) across trials within each of the four agent categories.

\emph{Participants.}
Twenty Tufts University undergraduate and graduate students participated.
Due to equipment failure, data were unavailable for two participants.
Thus, $18$ participants ($8$ male) with ages ranging from 18 to 30 years ($M$$=$$20.24$, $SD$$=$$3.61$) were included in our final sample.

\textbf{Results}\\
To confirm the assumptions of our study design and test our hypothesis, a repeated-measures ANOVA was conducted on each of the subjective ratings with \emph{agent category} as the independent variable.
For each ANOVA, the assumption of equal variance was confirmed using Mauchly's test of sphericity or otherwise adjusted.
In cases of violation, the degrees of freedom and corresponding $p$-value reflect either a Greenhouse-Geisser or Huynh-Feldt adjustment as per \cite{Girden1992}.
In addition, all post-hoc contrasts reflect a Bonferroni-Holm correction for multiple comparisons.

\emph{Human Similarity.}
We first tested participants' \emph{human similarity} ratings to determine whether our four-level manipulation of agent appearance elicited different attributions.
Specifically, we assumed the four levels would be perceived as having increasing human likeness from \emph{low} (lowest) to \emph{human} (highest).
As expected, the ANOVA showed a main effect of \emph{agent category} on \emph{human similarity} ratings: $F(3,51)$$=$$148.47$, $p$$<$$.01$, $\eta^{2}$$=$$.90$.
Furthermore all pairwise comparisons were significant ($p$$<$$.01$), confirming that participants' perceptions of the agents' human similarity were consistent with those assumed -- increasing from robots categorized as \emph{low} ($M$$=$$1.98$, $SD$$=$$.79$) to \emph{high} ($M$$=$$5.35$, $SD$$=$$1.70$) in human similarity to \emph{human} ($M$$=$$8.69$, $SD$$=$$.68$).

\emph{Eeriness.}
Based on Mori's uncanny valley theory, we assumed the four categories would elicit differentially negative evaluations, with the greatest eeriness attributed to highly humanlike robots and least to humans.
As expected, the ANOVA showed a main effect of \emph{agent category} on \emph{eeriness} ratings: $F(3,51)$$=$$52.99$, $p$$<$$.01$, $\eta^{2}$$=$$.76$.
Again, all pairwise contrasts were significant and consistent with assumptions.
Specifically, the highly humanlike robots were rated as most eerie ($M$$=$$6.52$, $SD$$=$$.85$) and \emph{humans} as least ($M$$=$$3.07$, $SD$$=$$.99$).
See Figure~\ref{fig:eeriness} (left).

\section{Study 2}
The findings of Study 1 suggested the existence of an uncanny valley with real-world robots.
To determine the \emph{impact} of the valley on human-robot interactons, here in Study 2 we tested whether it provokes avoidant behavior in observers \cite{STRAIT2015}.
Specifically, via a slight modification of the picture-viewing protocol, we tested whether highly humanlike robots are so emotionally motivating that participants attempt to end their encounters more frequently than those with less humanlike and human agents (\emph{Hypothesis 2}).

\begin{marginfigure}[0pc]
\begin{minipage}{\marginparwidth}
\centering
\includegraphics[width=1\marginparwidth]{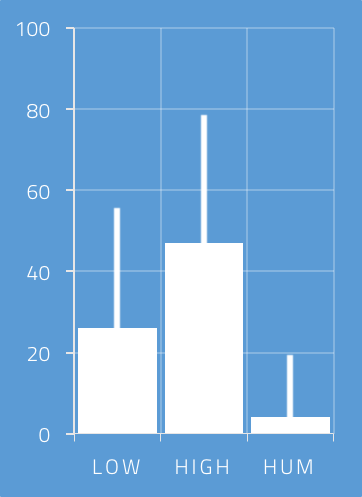}
\includegraphics[width=1\marginparwidth]{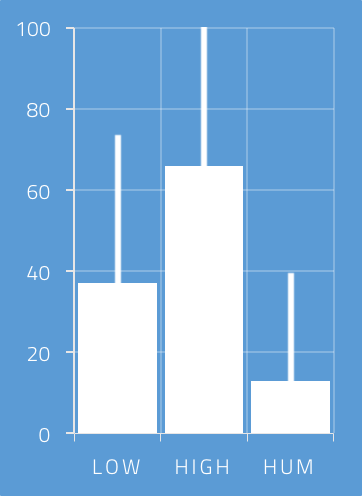}
\caption{Percentage of encounters terminated due to being unnerved, by agent category, in Studies 2 (top) and 3 (bottom).}
\label{fig:presses}
\end{minipage}
\end{marginfigure}

\textbf{Materials \& Methods}\\
\emph{Protocol.}
To investigate whether the appearance of highly humanlike robots is so distressing that people avoid their encounters, we adapted the protocol by Vujovi\'{c} and colleagues for studying aversive responding towards negative stimuli \cite{VUJOVIC2014}.
Here, the above set of $60$ images were each presented for up to $12s$.
Participants were informed that, should they wish to do so, \emph{they could press the spacebar to remove the given image from the screen}.
If the subject did \emph{not} press the spacebar, the image remained on display for a total viewing duration of $12s$.
Following the viewing, participants were cued to select one of three reasons for either pressing (\emph{unnerved}, \emph{bored}, or \emph{other}) or not pressing (\emph{interested}, \emph{indifferent}, or \emph{other}) the spacebar.
The choice of these options served to tease apart whether a stimulus creates a negative situation for the subject (being unnerved) or rather, whether a press response indicates some other motivating factor (e.g., boredom).
After a response was recorded, participants were then prompted to rate the appearance of the given agent as done in Study 1.

\emph{Measures.}
We again collected subjective ratings of the agents' \emph{human similarity} and \emph{eeriness}.
To index avoidant behavior, we recorded the percentage of button presses to measure the frequency of attempts to end encounters with the various agents, as well as participants' explicit reasons as to why they did or did not press the button.\footnote{Due to space constraints, only one measure of avoidant behavior -- press frequency due to being unnerved -- will be discussed. Note however, the other results are available in \cite{STRAIT2015}.}

\emph{Participants.}
Sixty-two undergraduates participated.
Due to equipment failure data were unavailable for two subjects, thus sixty subjects ($28$ male) with ages ranging from 18 to 28 years ($M$$=$$19.13$, $SD$$=$$1.48$) were included in our final sample.

\textbf{Results}\\
As in Study 1, a repeated-measures ANOVA was conducted on each of the dependent measures with \emph{agent category} as the independent variable.
Similarly, corrections (e.g., Bonferroni-Holm) were applied as appropriate.

\emph{Human Similarity.}
As in Study 1, the ANOVA on \emph{human similarity} ratings showed a main effect of agent category: $F(2.50,147.94)$$=$$788.85$, $p$$<$$.01$, $\eta^{2}$$=$$.93$.
All pairwise comparisons were significant, confirming that participants' perceptions of the agents' human similarity were consistent with those assumed -- increasing from robots categorized as \emph{low} ($M$$=$$2.44$, $SD$$=$$1.10$) to \emph{high} ($M$$=$$6.46$, $SD$$=$$1.38$) to \emph{human} ($M$$=$$8.90$, $SD$$=$$.27$.

\emph{Eeriness.}
Similarly, the ANOVA on \emph{eeriness} ratings also showed a main effect of agent category:\\
$F(2.47,145.74)$$=$$219.31$, $p$$<$$.01$, $\eta^{2}$$=$$.79$.
Again, all pairwise contrasts were significant and consistent with assumptions.
Specifically, the highly humanlike robots were rated as most eerie ($M$$=$$5.30$, $SD$$=$$1.39$) and \emph{humans} as least ($M$$=$$1.50$, $SD$$=$$.74$).
See Figure~\ref{fig:eeriness} (middle).

\emph{Avoidant Behavior.}
The ANOVA on \emph{press frequency} -- the frequency at which participants terminated their encounters -- also showed a main effect of agent category:\\
$F(2.50,102.55)$$=$$36.18$, $p$$<$$.01$, $\eta^{2}$$=$$.46$.
Specifically, participants terminated their encounters with highly humanlike robots more frequently and due to being unnerved ($M$$=$$.47$, $SD$$=$$.31$) relative to less humanlike robots ($M$$=$$.26$, $SD$$=$$.29$) and humans ($M$$=$$.04$, $SD$$=$$.15$).
See Figure~\ref{fig:presses} (top).

\section{Study 3}
Study 2 made two theoretical contributions.
First, it confirmed the existence of an uncanny valley in humanoid robots in replicating the findings of Study 1.
Second, it established support towards Mori's speculations that highly humanlike entities can be so unnerving that they motivate avoidant behavior.
However, there remains a key limitation -- both within Study 2 and across uncanny valley literature at large.
Specifically, the question of a person's exposure to and experience with social robots remains an unaddressed critique (e.g., \cite{HANSON2005}).
Thus, we developed Study 2 as a follow-up investigation.
The primary goal was to determine whether controlled exposure to a humanoid would attenuate aversive responding towards robots in general and whether it would extinguish the presence of an uncanny valley or any valley effects in particular.
This follow-up study thus allowed us to conceptually replicate the findings in a context in which key limitations of Study 2 were resolved.

\textbf{Materials \& Methods}\\
\emph{Protocol.}
Here, participants were preexposed to one of three agents (either a robot with low or moderate human similarity, or a human confederate) in a simple interactive task (developed in \cite{STRAIT2014}) prior to completing the picture-viewing task employed in Study 2.

\emph{Participants.}
Seventy-one undergraduate and graduate students participated ($29$ male), ranging from $18$ to $36$ years old ($M=20.27$, $SD=2.94$).

\textbf{Results}\\
\emph{Human Similarity.}
As in Studies 1--2, the ANOVA on \emph{human similarity} ratings showed a main effect of agent category: $F(2,136)$$=$$741.72$, $p$$<$$.01$, $\eta^{2}$$=$$.92$.
All pairwise comparisons were significant, confirming that participants' perceptions of the agents' human similarity were consistent with those assumed -- increasing from robots categorized as \emph{low} ($M$$=$$3.02$, $SD$$=$$1.20$) to \emph{high} ($M$$=$$5.05$, $SD$$=$$1.30$) to \emph{human} ($M$$=$$8.77$, $SD$$=$$.47$).

\emph{Eeriness.}
Similarly, the ANOVA on \emph{eeriness} ratings also showed a main effect of agent category:\\ $F(1.76,119.54)$$=$$219.95$, $p$$<$$.01$, $\eta^{2}$$=$$.76$.
Again, all pairwise contrasts were significant and consistent with assumptions.
Specifically, the highly humanlike robots were rated as most eerie ($M$$=$$6.29$, $SD$$=$$1.23$) and \emph{humans} as least ($M$$=$$2.87$, $SD$$=$$1.51$).
See Figure~\ref{fig:eeriness} (right).

\emph{Avoidant Behavior.}
The ANOVA on \emph{press frequency} -- the frequency at which participants terminated their encounters -- also showed a main effect of agent category:\\
$F(2,74)$$=$$35.06$, $p$$<$$.01$, $\eta^{2}$$=$$.49$.
Specifically, participants terminated their encounters with highly humanlike robots more frequently and due to being unnerved ($M$$=$$.66$, $SD$$=$$.34$) relative to less humanlike robots ($M$$=$$.37$, $SD$$=$$.36$) and humans ($M$$=$$.13$, $SD$$=$$.26$).
See Figure~\ref{fig:presses} (top).

\section{General Discussion}
\emph{Summary of Present Research.}
Studies 1--3 present an experimental test of Mori's uncanny valley theory, as it pertains to real-world robots of varying human similarity and humans.
In Study 1, we measured people's subjective ratings of the agents' eeriness, which reflected a valley corresponding to robots that are highly humanlike in their appearance.
Study 2 replicated this valley in eeriness ratings and demonstrated the use of a novel protocol for more objective measurement of valley effects.
The results showed that not only do people rate highly humanlike robots as more eerie, but moreover, they exhibit greater avoidance of such encounters than those with less humanlike and human agents.
In Study 3, despite preexposure to an embodied humanoid prior to the picture-viewing protocol, the valley in participants' ratings of eeriness and their corresponding avoidance of highly humanlike robots persisted.
Consistent with Mori's original postulations, these findings robustly demonstrate that robots can be so unnerving that they motivate people to avoid them.
Furthermore, they suggest that people's aversion to highly humanlike robots is not sensitive to other social factors such as exposure.

\emph{Future Impact.}
The present work has clear theoretical and practical implications.
Studies 1--3 both demonstrate and replicate an uncanny valley in the appearance of humanlike robots and people's avoidance thereof.
In doing so, they provide strong support of Mori's original theory theory and moreover, establish that robust valley effects pose a formidable design consideration for human-robot interaction.
Should these findings replicate with other anthropomorphized technologies such as user interfaces equipped with humanlike attributes, it will show the uncanny valley extends beyond robotics to pose a broader design consideration for human-computer interaction at large.

The long-term impact of this work is three-fold.
First, with the establishment of clear and robust effects, the present work allows the community to now focus on the causal mechanisms by which certain humanlike appearances contribute to people's discomfort.
It remains to be explained as to \emph{why} people are particularly averse to highly humanlike robots.
Numerous plausible explanations have been proposed, but there have been inconsistent findings due to the variable nature of subjective assessment.
This highlights a second contribution: the protocol described here has the potential to address outstanding inconsistencies via a more objective measurement of valley effects (observation of people's behavior and their emotional experiences driving their actions).
Last but not least, further application of knowledge gained from the present work and the demonstrated protocol may serve towards the design of future robots.
For example, in iterative design practices, the protocol may be used as a quick and simple test of the efficacy of one design versus others.

\section{Acknowledgements}
I am grateful to my advisor, Heather Urry, for her tremendous support and significant advising towards the work presented here.
I would also like to thank Victoria Floerke and Lara Vujovi\'{c} for their help and collaboration on Study 1, as well as for their support and advice during the Study 2 production.
Lastly, I would like to thank the many undergraduate research assistants in the Emotion, Brain, and Behavior Laboratory, as well as Max Bennett, George Brown, Brendan Fleig-Goldstein, and Maretta Morovitz who contributed towards data collection, processing, and analysis.

\balance{}

\bibliographystyle{acm-sigchi}
\bibliography{sample}

\end{document}